# Self-guiding of long-wave infrared laser pulses mediated by avalanche ionization

D. Woodbury[1], A. Goffin[1], R. M. Schwartz[1], J. Isaacs[2], and H. M. Milchberg[1],*

[1]Institute for Research in Electronics and Applied Physics, University of Maryland, College Park, MD 20742, USA
[2]Plasma Physics Division, U.S. Naval Research Laboratory, 4555 Overlook Ave. SW, Washington, DC 20375, USA

**Abstract:** Nonlinear self-guided propagation of intense long-wave infrared (LWIR) laser pulses is of significant recent interest owing to the high critical power for self-focusing collapse at long wavelengths. This promises transmission of very high power in a single filament as opposed to beam breakup and multi-filamentation. Here, using the most current picture of LWIR ionization processes in air, we present extensive simulations showing that isolated avalanche sites centered on aerosols can arrest self-focusing, providing a route to self-guided propagation of moderate intensity LWIR pulses in outdoor environments.

Femtosecond filamentation of intense laser pulses in gases and condensed media arises from the interplay of diffraction, Kerr self-focusing, and collapse arrest by plasma-induced refraction, enabling high intensity self-guided propagation over extended distances [1]. Filamentation occurs for pulses whose peak power exceeds a critical value $P_{cr} = 3.77\lambda^2/8\pi n_0 n_2$ for Gaussian beams, where $\lambda$ is the laser wavelength, and $n_0$ and $n_2$ are the medium's linear and nonlinear indices of refraction. In "standard" filamentation, self-induced Kerr lensing focuses the beam until multi-photon or tunneling ionization of the medium and associated plasma defocusing arrests pulse collapse. As input power is increased well beyond $P_{cr}$, the beam is unstable to breakup into multiple filaments, limiting the peak power delivered in a single high intensity channel. The $P_{cr} \propto \lambda^2$ scaling indicates higher multi-filamentation thresholds for longer wavelengths, stimulating recent interest in mid-IR and long-wave IR (LWIR) filamentation [2-9]. For LWIR pulses, new mechanisms have been proposed for collapse arrest, including the formation of optical shocks and harmonic walk-off for short (<1 ps) pulses [5-7] and avalanche ionization seeded by many-body induced ionization for longer (>1 ps) pulses [8-10].

In this Letter, we present propagation simulations showing that avalanche ionization at discrete breakdown sites, likely seeded by aerosols, is essential for atmospheric self-guiding of moderate intensity LWIR pulses with few-millimeter beam widths, consistent with recent experiments [9]. Our simulations incorporate the latest understanding of LWIR ionization processes in air provided by our recent experiments [11-13]. In general, avalanches proceed as localized plasma breakdowns centered either on aerosols or electrons generated by tunneling ionization early in the pulse. In both cases, refraction from these discrete plasma sites is manifested through forward Mie scattering [14]. We incorporate these spatially discrete, transient breakdowns in our propagation simulations, a more realistic approach than the continuous plasma assumed in previous models. In the absence of aerosols, we find that self-focusing continues until it is arrested by standard tunneling ionization, with avalanche-generated plasma refracting only the trailing edge of the pulse. Aerosols, on the other hand, lead to enhanced, *saturable* ionization early in the pulse, enabling avalanche-mediated collapse arrest and channeling of few picosecond LWIR pulses at moderate intensities.

In avalanche breakdown, free electrons undergo laser-driven dephasing elastic collisions with neutral molecules until they have enough kinetic energy to collisionally ionize the neutrals in a cascading process. The growth in the local number of electrons at an avalanche site is $n_e = n_{eo} e^{\nu_i t}$ where $\nu_i = \langle \sigma_i v \rangle N_n$ is the electron collisional growth rate, $n_{eo}$ is the local number of seeds, $N_n$ is the local neutral molecule density, and $\langle \sigma_i v \rangle = \int_0^\infty dv\, f(v) \sigma_i(v) v$ for electron velocity distribution $f(v)$ and collisional ionization cross section $\sigma_i$. Growth saturates as the neutral density is depleted ($N_n = N_{no} - N_e$) for increasing electron density $N_e$.

At an avalanche site, electron diffusion gives a characteristic plasma radius $r_d = \sqrt{2\tau k_B T_e / m_e \nu_{en}} \sim 0.3\sqrt{\tau[\text{ps}]\, T_e\,[\text{eV}]}$ µm for electron temperature $T_e$ and electron-neutral collision rate $\nu_{en} \sim 2-4$ ps$^{-1}$ (for 2 eV $< T_e <$ 1 keV) [15,16]. In particular, diffusion is limited to ~5 µm even for $k_B T_e \sim 100$ eV for a <3.5 ps pulse, so we use $r_d \sim$ 5 µm as a baseline for our propagation simulations below. In general, breakdowns proceed as isolated avalanche sites surrounded by neutral air unless the seed electron density satisfies $N_{e0} > 1/r_d^3$ ~$10^9$–$10^{11}$ cm$^{-3}$, for which the laser effectively interacts with a continuous plasma.

During laser propagation, the absolute changes in refractive index ($\Delta n = |n-1| \sim \overline{N_e}/2N_{cr}$) and propagation phase ($|\Delta\Phi| = 4\pi a |\Delta n|/\lambda$) across an individual breakdown site of radius $a$ are small ($\ll 1$) for average plasma density well below critical density, $\overline{N_e} \ll N_{cr} = 1.1 \times 10^{21}$ cm$^{-3}/\lambda^2$ [μm]. Thus scattering occurs in the Rayleigh-Gans (RG) regime ($|\Delta n| \ll 1$, $a \ll \lambda/|\Delta n|$) [14]. An effective refractive index is calculated from the forward scattering amplitude [14] $S(0)$ of an ensemble of scatterers of number density $N_{sc}$, giving $n_{eff} = 1 + N_{sc} V \Delta n = 1 - N_{sc} V (\overline{N_e}/2N_{cr})$, for average breakdown site volume $V$ [17]. This equals the index of a continuous plasma of density $N_{sc} V \overline{N_e}$, as covered in previous work on exploding nano-plasmas [26]. The RG approximation breaks down as $\overline{N_e}$ approaches $N_{cr}$ at individual breakdown sites, necessitating Mie scattering calculations [14]. This limits our use of RG-based $n_{eff}$ to breakdowns with $\overline{N_e} < N_{cr}/2 \sim 5 \times 10^{18}$ cm$^{-3}$, corresponding to ~6×10$^9$ electrons in a $a = r_d = 5$ μm breakdown volume, for which the Mie and RG models give effective refractive indices within 20% [17]. Above this density, our simulations have limited fidelity.

Our goal is to couple a model of avalanche at discrete sites to a propagation simulation. The best, albeit forbidding, approach would be to solve the Boltzmann equation for the full time-resolved electron distribution function [27-30], accounting for angle-resolved scattering over wide primary and secondary electron energy ranges. Instead, we describe a simpler temperature-based model [31-33]. This model contains the key physics to explain recent LWIR propagation experiments, and is insensitive to our assumption of a Maxwellian electron energy distribution.

First we consider electron heating. A free electron in a laser field has a cycle-averaged kinetic energy $U_p \cong 0.93\, I$ [TW/cm$^2$]$(\lambda[\mu m])^2$ eV, which is transferred into incoherent motion through electron-neutral collisions at rate $\nu_{en}$. The Lorentz-Drude model gives a heating rate per electron of $W_{coll} = 2U_p \omega^2 \nu_{en}(\omega^2 + \nu_{en}^2)^{-1} \sim 2U_p \nu_{en}$, for laser frequencies $\omega/\nu_{en} \gg 1$ at atmospheric pressure. Electron heating is offset by losses: rovibrational and electronic excitation, dissociation and ionization losses of energy $\chi_l$ and excitation rate $\nu_l$ in N$_2$ and O$_2$ [15-16]. The rate of change in total system internal energy density is then

$$\frac{dU}{dt} = \frac{3}{2}\frac{d(k_B N_e T)}{dt} = \frac{3}{2}k_B\left(\frac{dN_e}{dt}T + \frac{dT}{dt}N_e\right) = 2\nu_{en} N_e U_p + \nu_i N_e U_p - \sum_l N_e \nu_l \chi_l, \quad (1)$$

where $k_B$ is Boltzmann's constant. Here, we have added the term $U_p(dN_e/dt) = \nu_i N_e U_p$, which accounts for the effective heating of electrons collisionally released during the laser cycle. We ignore diffusive losses for laser spot sizes much larger than $r_d$: temperature gradients are weak because $\nu_{en}$ greatly exceeds the electron-electron and electron-ion collision rates, with no spatial dependence until saturation at the end of the breakdown [17]. Rearranging gives

$$\frac{dk_B T}{dt} = \frac{2}{3}\left(2U_p \nu_{en} + \nu_i U_p - \Sigma_l \nu_l \chi_l\right) - \nu_i T \; , \tag{2}$$

where $N_e^{-1}(dN_e/dt)T = \nu_i T$ tracks thermal energy redistribution in a growing electron population [31-33]. Here, we directly integrated relevant cross sections in $N_2$ and $O_2$ [15-16] over a Maxwellian distribution up to $k_B T$=1 keV instead of relying on low temperature (<30 eV) tabulated rates [31-34]. The heating rate $2U_p(2\nu_{en} + \nu_i)/3$ and loss rate $2(\Sigma_l \nu_l \chi_l)/3 + \nu_i T$ in Eq. (2) are plotted vs. $k_B T$ on the left scale in Fig. 1 for a λ=10.2 μm, 1 TW/cm² pulse; the point where they cross ($dk_B T/dt = 0$) defines a quasi-static equilibrium temperature $T_s$ achieved for long drive pulses (after delay $\tau \gg \nu_{en}^{-1}$) [11,32] and an associated quasi-static growth rate $\nu_{is}$. The right scale shows the ionization rate $\nu_i(k_B T)$ as a function of temperature and, for comparison, the rate $\nu_i(E)$ for a monoenergetic electron distribution.

Figure 1(b) plots the calculated quasi-static growth rate $\nu_{is}$ (for $\tau \gg \nu_{en}^{-1}$) as a function of intensity, compared with limiting cases and other theoretical and experimental results. The horizontal dashed line indicates the maximum achievable growth rate matching the peak ionization rate in Fig. 1(a). The other two dashed curves indicate growth rates assuming that laser heating immediately results in ionization of species with ionization potential $\chi_p$ (upper (yellow) dashed curve, $\nu_i = 2U_p \nu_{en}/\chi_p$) or assuming electron velocities are given solely by quiver motion in the laser field, $v_q(t) = (eE/m\omega)\cos(\omega t)$ (lower dashed curve, appropriate for $\tau \ll \nu_{en}^{-1}$ [35]). These limits define two regimes: above 5 TW/cm², $\nu_{is}$ is insensitive to details of the model, since it is tightly bounded by the limiting cases. At lower intensities (1-5 TW/cm²), $\nu_{is}$ is within a factor of ~10× of the limiting cases. Absent full Boltzmann calculations, we scale [17] the effective field from DC breakdown modeling in $N_2$ [27], as well growth rates from DC experiments in $N_2$ [27,36] and λ=4 μm experiments in air [12,13], all showing good agreement with $\nu_{is}$. Computing our model's rate coefficients using non-Maxwellian electron distributions or including inelastic contributions to collisional heating give similar variation (~2×) in $\nu_{is}$ [17]. Thus, using a Maxwellian distribution should still allow accurate discrimination between high intensity (>10 TW/cm$^2$) and low intensity (0.5-5 TW/cm$^2$) regimes, while also incorporating the delayed response [37] of avalanche growth to intensity transients.

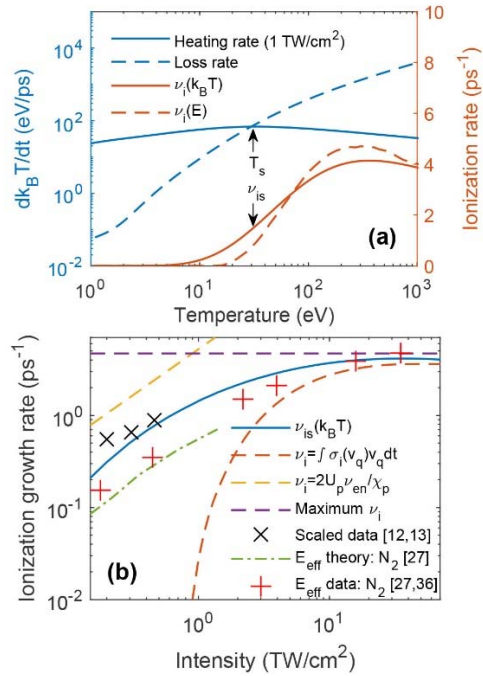

**Fig. 1.** (a) Temperature-dependent heating rate (for λ=10.2 μm, 1 TW/cm$^2$) $(dk_BT/dt)_{heating} = 2U_p(2\nu_{en} + \nu_i)/3)$ and loss rate $|(dk_BT/dt)_{loss}| = 2(\Sigma_l\nu_l\chi_l)/3 + \nu_iT$ (left scale, log) in air. The ionization rate $\nu_i$ (right scale) is shown as a function of electron energy $E$ and temperature $T$. (b) Quasi-static ionization growth rate $\nu_{is}$ extracted from the temperature model (solid blue curve). Dashed curves indicate limiting values of the growth rate based on no collisional heating (red), no energy loss (yellow), and the peak value of the ionization growth rate (purple). Scaled rates from Boltzmann theory for DC breakdowns in $N_2$ (dash-dot) and from experimental results in air at λ=4 μm (×) and DC experiments in $N_2$ (+) are given for comparison.

The initial electron population needed to seed LWIR avalanche in air can originate from tunneling ionization, which depends extremely sensitively on intensity as seen in Fig. 2. A ubiquitous air contaminant with $\chi_p$~6 eV, as recently measured at relative concentrations

~$10^{-9}$–$10^{-11}$ [13], dominates ionization below ~10 TW/cm². Separately, aerosols (solid density particulates including dust, water droplets/fog, etc.) are readily ionized due to near-field enhancement or existing static charge. We recently estimated aerosol concentrations of ~$10^4$ cm$^{-3}$ in our lab air by measuring the number of avalanches inside a breakdown threshold volume with and without particulate filtering [12]. Other detailed measurements of aerosols in indoor environments show a range of number concentration ($N_{sc}$~$10^2$–$10^4$ cm$^{-3}$) and particle size (~0.1-10 μm) [38-40]. In general, aerosol concentrations are higher in outdoor "field" conditions envisioned for applications of self-guiding.

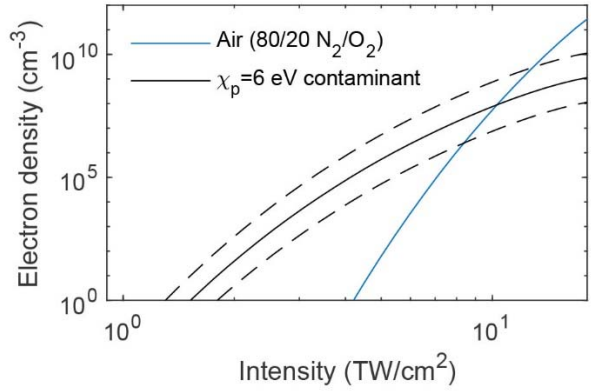

**Fig. 2.** Generated electron density vs. peak intensity from ionization of air and a $\chi_p$~ 6 eV contaminant [13] by a λ=10.2 μm, 1 ps FWHM Gaussian pulse. Dashed lines indicate bounds of approximate uncertainty in the number density of contaminant species [13].

We can estimate the requirements for collapse arrest and self-guiding by equating the nonlinear index shifts associated with Kerr focusing and plasma refraction: $\Delta n_{Kerr} = \Delta n_{eff} \rightarrow n_2 I = N_{sc} V \overline{N_e}/2N_{cr}$. For $n_2$ ~$5 \times 10^{-19}$ cm²/W [41,42] and $V = 4\pi r_d^3/3$, we estimate $(N_{sc} V \overline{N_e})$~$10^{13}\times I$[TW/cm²] cm$^{-3}$. At $I = 1$ TW/cm², this gives $N_{sc}$~$10^4$ cm$^{-3}$ breakdown sites avalanched to $\overline{N_e}$~ $N_{cr}/2$, the limit of our effective index approximation, or a larger $N_{sc}$ avalanched to a lower terminal $\overline{N_e}$, or a continuum density $N_e = N_{sc} V \overline{N_e}$.

Propagation simulations were conducted by combining the avalanche model ($n_e = n_{eo} e^{v_i t}$ plus Eq. (2)) with a 2D+time axisymmetric unidirectional pulse propagation equation (UPPE) solver [43,44]. Our simulation parameters match a recent experiment with 3.5 J, 3.5 ps, λ=10.2 μm pulses ($P$~$2P_{cr}$) focused to a 4 mm FWHM spot (4 TW/cm², ~5 m Rayleigh range) [9]. The pulses initially self-focused and created a tenuous visible plasma over ~5 m, followed by beam expansion to ~1 cm FWHM and self-guiding over ~30 m at peak intensities ~1 TW/cm², accompanied by pulse shortening to ~1.8 ps. Since generation of seed electrons from tunneling ionization is negligible at this intensity, the avalanche-generated plasma responsible for self-guiding was thought to be seeded by many-body induced ionization [8-10]. Our recent work [13] cast doubt on this seed source and pointed instead to tunneling ionization of contaminants or aerosols.

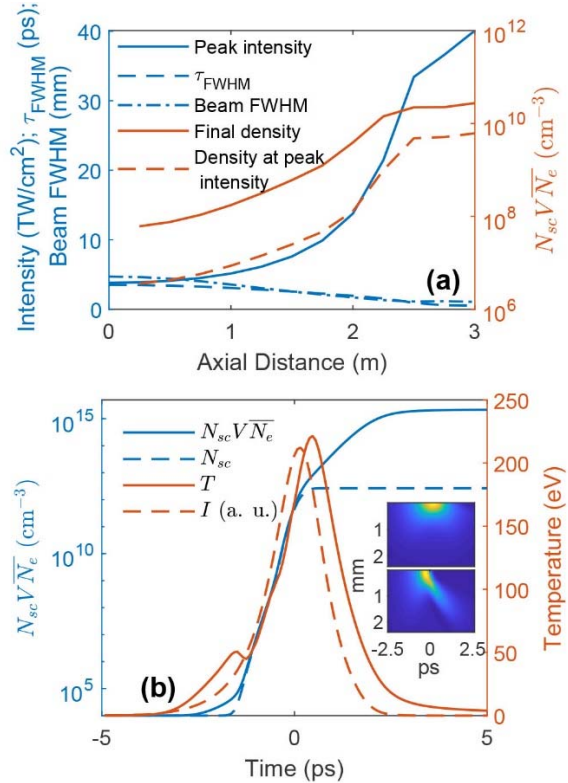

**Fig. 3.** (a) Pulse parameters over 3 m of propagation. Left axis: pulse peak intensity (TW/cm², solid) and beam FWHM and temporal FWHM (mm, ps; dashed), right axis, density at the intensity peak of the pulse (dashed) and after the pulse has completely passed (solid). (b) On-axis intensity and plasma density/temperature after 2.25 m of propagation, showing a rapid increase in the density $N_{sc}$ of breakdown sites due to seed generation by tunneling (dotted blue line), followed by slower increase in volume average density $N_e = N_{sc} V \overline{N_e}$ due to avalanche. Temperature (right scale) roughly follows the pulse intensity profile. Inset plots show spatiotemporal intensity profiles (normalized to peak intensity) at 2.25 m (top) and 3 m (bottom).

We first consider aerosol-free air, initializing propagation at the focus (4 mm FWHM, 4 TW/cm²). The initial electron population was set at $N_{e0} = 10^4$ cm$^{-3}$, but results were insensitive to this value. Additional seed electrons were contributed by tunneling ionization [45] (see Fig. 2). Figure 3(a) shows the peak intensity, pulse temporal FWHM and beam FWHM, and plasma density (at the peak intensity and after the pulse) versus propagation distance. The pulse self-focuses to ~20 TW/cm² after 2.25 m of propagation while also undergoing self-shortening to ~2 ps as the front and rear diffract. At this point, substantial tunneling ionization occurs during the leading edge of the pulse, generating a seed electron density $N_{sc} \sim 10^{12}$ cm$^{-3} > r_d^{-3}$, in the continuum density regime. This is seen in Fig. 3(b), which shows the on-axis intensity and plasma density after 2.25 m of propagation. The falling edge of the pulse drives continued avalanche to high density behind the pulse, as effective density $N_e = N_{sc}\sqrt{N_e}$ continues growing above the density $N_{sc}$ of sites generated through tunneling. This rising density further refracts the pulse's falling edge, leading to the shortened pulses shown in the inset plots of Fig. 3(b) at 2.25 m and 3 m. The final sub-picosecond pulse continues to filament in the tunneling ionization regime, with peak intensity $I \gtrsim 20$ TW/cm² and ~1 mm FWHM beam size [4,7], very different than the ~1 TW/cm², 1 cm beam observed in [9]. Since our envelope-resolved simulation does not have sufficient spatial resolution to track further propagation with high fidelity, we terminated this simulation at ~3 m.

The failure of avalanche growth in aerosol-free air to arrest collapse before the onset of tunneling is inevitable given the electron density growth rates in Fig. 1(b) and tunneling yields in Fig. 2. Even for the maximum growth rate of 4.7 ps$^{-1}$ (predicted only for $I \gtrsim 10$ TW/cm² in Fig. 1(b)), an initial electron density $N_{eo} \sim 10^9$ is needed to reach the effective density $N_{sc}\sqrt{N_e} = 10^{13}$ cm$^{-3}$ we estimated for self-guiding at 1 TW/cm². This initial density in turn requires tunneling at ~10 TW/cm² in the leading edge of the pulse as shown in Fig. 2. Thus, it is unlikely that the ~1 TW/cm² channeling observed in [9] is stabilized by avalanche from electrons liberated by tunnel ionization.

Avalanche ionization in aerosols, however, introduces a new propagation regime consistent with self-guiding at modest intensity. For the near-solid density of an aerosol particle, the electron-neutral collision rate ($\nu_{en} \sim 10^{15}$ s$^{-1}$) is much higher than infrared frequencies ($\omega_{en}/\nu_{en} \ll 1$) [28,46]. Thus collisional heating $W_{coll} \propto I\nu_{en}^{-1}(1 + (\omega_{en}/\nu_{en})^2)^{-1} \approx I/\nu_{en}$ becomes wavelength-independent for $\lambda \gtrsim 300$ nm, suggesting ionization rates calculated for breakdown in fused silica by λ=1 μm pulses are representative [28]. Using these growth rates, a 3.5 ps, 1 TW/cm² pulse fully singly ionizes all atoms in a ~0.2 μm radius particle in the leading edge of the pulse, after accounting for field enhancement at the particle surface [17]. This high-density plasma would then explode into the surrounding air and continue to avalanche.

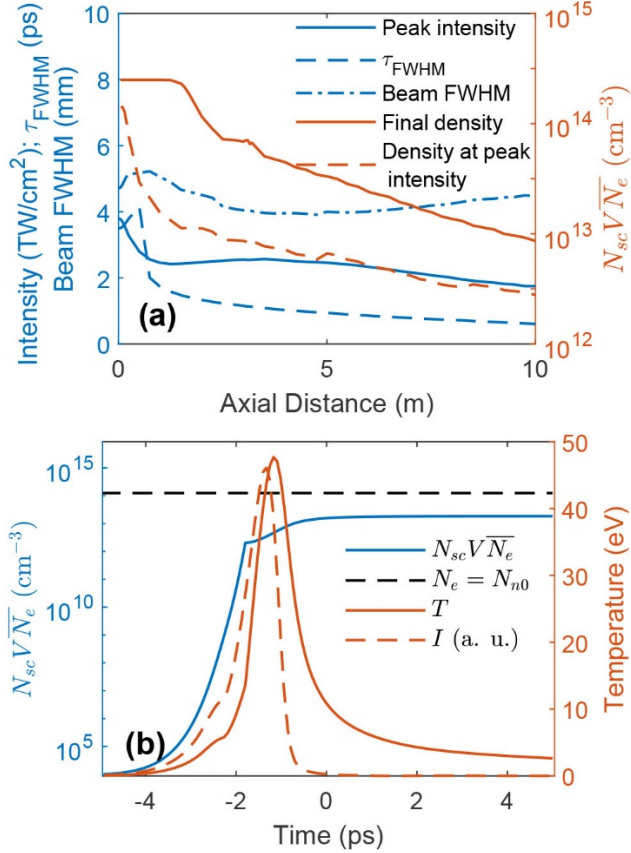

**Fig. 4.** (a) Simulation including aerosol-enhanced avalanche for initial aerosol density of $2\times10^4$ cm$^{-3}$. (b) On-axis intensity and plasma density/temperature after 6 m of propagation, showing a rapid increase in density due to breakdowns in aerosols, followed by slower avalanche in air, with volume average density reaching ~$10^{13}$ cm$^{-3}$ near the intensity peak of the pulse, broadly consistent with self-guiding as shown above. The horizontal dashed line indicates full ionization of the breakdown sites.

Accordingly, we conduct a second set of propagation simulations for a range of initial aerosol densities ($N_{sc}\sim 10^3 – 3\times 10^4$ cm$^{-3}$), with results for $N_{sc} = 2 \times 10^4$ cm$^{-3}$ shown in Fig. 4. For simplicity and specificity, we assume a uniform aerosol radius of $a = 0.2$ μm. In these simulations, the pulse is immediately shortened as energy in the falling edge is scattered off-axis by the rapidly formed plasmas. After this transient, aerosol-enhanced plasma generation is sufficient to arrest self-focusing, resulting in self-guiding of $0.5 − 1.5$ ps pulses at ~2-2.5 TW/cm$^2$ over ~10 m. Runs for lower aerosol densities ($N_{sc} < 10^4$ cm$^{-3}$) showed slower self-focusing, with collapse ultimately arrested by tunneling ionization [17], leading to behavior similar to Fig. 3.

Figure 4(b) shows on-axis plasma density and pulse shape at 6 m, after onset of self-guiding. In contrast to Fig. 3, the aerosols provide a rapid increase in $N_{sc}V\overline{N_e}$ that arrests collapse over the full pulse. Ionization saturation limits further pulse shortening, which would reduce avalanche for $\tau < \nu_{en}^{-1}$. Thus avalanche breakdown of aerosols appears essential for long distance stable self-guiding of moderate intensity picosecond LWIR pulses.

While our simplifications have made a complex problem tractable and provided a clear physical picture for the role of aerosols, our work calls for more detailed studies. In particular, because local density reaches $\overline{N_e} \geq N_{cr}/2$ at breakdown sites during the pulse, and because the saturated aerosol plasma can further heat through electron-ion collisions and expand into the background air, the RG model breaks down and would need to be replaced by a full scattering computation coupled to plasma dynamics.

In summary, we have developed a model for self-guiding of long wavelength infrared (LWIR) picosecond laser pulses stabilized by avalanche ionization from discrete plasma breakdown sites. In aerosol-free air, we find pulse self-focusing is arrested by tunneling ionization, leading to high self-guided intensities and narrow channel diameters, inconsistent with recent experiments [9]. Aerosol-centered avalanche sites, however, enhance plasma generation, which enables self-guiding at moderate intensities with larger channel diameters. Future experimental propagation studies of picosecond LWIR pulses in controlled atmospheres could help quantify these effects.

*Acknowledgements.* The authors thank P. Sprangle, M. Kolesik, J. Moloney, S. Tochitsky, and C. Joshi for useful discussions, and J. K. Wahlstrand for help calculating tunneling ionization rates. This work is supported by the Office of Naval Research (ONR) (N00014-17-1-2705) and the Air Force Office of Scientific Research (AFOSR) (FA9550-16-1-0121, FA9550-16-1-0284). D. W. acknowledges support from the DOE NNSA SSGF program under DE-NA0003864.

*Corresponding author: milch@umd.edu

# Supplementary material for "Self-guiding of long-wave infrared laser pulses mediated by avalanche ionization"

## 1. Effective refractive index under Rayleigh-Gans scattering

Here we describe the approach for finding an effective refractive index of a medium composed of many discrete breakdown sites, closely following the approach of [1]. For a scatterer located at the origin in spherical coordinates, a complex scattering function $S(\theta, \phi)$ is defined by $u_{scatt} = S(\theta, \phi)(e^{ik(r-z)}/ikr)u_0$, assuming that a scalar field description is appropriate. Here $u_{scatt}$ is the scattered field amplitude at $(r, \theta, \phi)$, $u_0 e^{ikz}$ is a plane wave incident on the scatterer, and $k = 2\pi/\lambda$ is the vacuum wavenumber. For an ensemble of scatterers of uniform size, all forward scattered light (described by $S(\theta = 0, \phi) \equiv S(0)$, independent of $\phi$) is coherent with the incident beam, yielding an effective refractive index

$$n_{eff} = 1 + \frac{2\pi N_{sc}}{k^3} \Im\{S(0)\}, \qquad (S.1)$$

where $N_{sc}$ is the number density of scattering particles and $\Im\{S(0)\}$ denotes the imaginary part of $S(0)$. Scattering into angles $\theta \neq 0$ is incoherent for random placement of scatterers, and contributes only to beam propagation losses, as covered below.

As discussed in the main text, diffusion during avalanche for 3.5 ps pulses dictates localized breakdown plasma radii $r_d \sim 5\mu m$ for $T_e \sim 100$ eV. The average plasma density at the local breakdown site is well below the critical density, $\overline{N_e} \ll N_{cr} = 1.1 \times 10^{21}$ cm$^{-3}/\lambda^2$ [μm] so that the phase shift for a beam traversing the plasma site is $|\Delta\Phi| = 2kr_d|n-1| \ll 1$, where $n = \sqrt{1 - N_e/N_{cr}} \approx 1 - N_e/2N_{cr}$ is the plasma index, $|n-1| = |\Delta n| = N_e/2N_{cr} \ll 1$, and $r_d \ll \lambda/|\Delta n|$ for $\lambda \sim 10$ μm. The two latter conditions ensure that scattering falls in the Rayleigh-Gans (RG) regime, a simple extension of the familiar Rayleigh approximation. In particular, $S(0)$ is the same for RG and Rayleigh scattering, so we readily obtain

$$S(0) = \frac{ik^3(n-1)}{2\pi} V, \qquad (S.2)$$

where $V$ is the volume of the scattering site. Combining Eqs. S.1 and S.2 we find

$$n_{eff} = 1 + \frac{2\pi N_{sc}}{k^3} \Im\left\{\frac{ik^3(n-1)}{2\pi} V\right\} = 1 - N_{sc}V(n-1) \approx 1 - N_{sc}V(\overline{N_e}/2N_{cr}). \qquad (S.3)$$

Since $N_{sc}V$ corresponds to an effective fill fraction, and $n - 1 \approx N_e/2N_{cr}$ for plasma well below the critical density, this result is simply the same as for an effective continuous plasma density $N_{e,eff} = N_{sc}V\overline{N_e}$.

## 2. Extinction (scattering) losses

The total $\theta \neq 0$ Mie scattering of the incident beam results in an extinction (attenuation) length given approximately by

$$\gamma_{ext} \sim N_{sc} \pi r_d^2 |n-1|^2 \approx N_{sc} \pi r_d^2 \frac{(\overline{N_e})^2}{2 N_{cr}^2} \ll 1. \tag{S.4}$$

For $r_d \approx 5\,\mu m$, $\lambda = 10\,\mu m$, and $\overline{N_e} \sim 5 \times 10^{18}\,cm^{-3}$ ($= N_{cr}/2$, the approximate limit of RG applicability) we get

$$\gamma_{ext} \approx N_{sc}\,[cm^{-3}] \times 10^{-7}\,cm^{-1}, \tag{S.5}$$

corresponding to an extinction length $\gamma_{ext}^{-1} \sim 10$ m for $N_{sc} \sim 10^4\,cm^{-3}$. For plasma site densities well below this, the $(\overline{N_e})^2$ dependence indicates that Mie scattering losses are negligible over the length scales we consider. Indeed, the lineouts in Fig 4(b) of the main text indicate that for the self-guided pulse interacting with aerosol initiated plasmas, $\overline{N_e} \sim N_{cr}/50$ at the intensity peak of the pulse, indicating $\gamma_{ext}^{-1} \sim 3$ km for $N_{sc} \sim 2 \times 10^4\,cm^{-3}$.

Once the breakdown fully singly ionizes, the scattering efficiency approaches a limiting value [1] such that for $r_d \approx 5\,\mu m$,

$$\gamma_{ext} \approx 2 N_{sc}\,\pi\,r_d^2 \approx N_{sc}\,[cm^{-3}] \times 10^{-6}\,cm^{-1}, \tag{S.6}$$

giving an extinction length $\gamma_{ext}^{-1} \sim 1$ m for $N_{sc} \sim 10^4\,cm^{-3}$. This extinction length will be even shorter as plasma breakdowns expand radially with additional laser heating. We note that since this is not absorption, but rather side and small angle scattering, laser energy may appear to persist beyond $\gamma_{ext}^{-1}$ as significant energy is scattered into an incoherent, forward directed "halo." In particular, the experiment [2], with which we compare our simulations, had a secondary post-pulse 25 ps after the main 3.5 ps pulse, which did not persist beyond the ~5 m plasma channel. The creation of near-critical density plasma sites from breakdown saturation of aerosols by the main pulse is consistent with (1) further heating and plasma emission from the second pulse to create a visible plasma channel, and (2) enhanced extinction of this second pulse over few meter lengths.

## 3. Transition to Mie regime

As the average plasma density $\overline{N_e}$ for a breakdown site approaches $N_{cr}$, the RG approximation breaks down and the index diverges from $n_{eff} = 1 - N_{sc} V (\overline{N_e}/2N_{cr})$. A full calculation of the Mie scattering amplitude for $\lambda = 10\,\mu m$ shows $n_{eff}^{Mie}$ agrees with $n_{eff}$ to within 20% for a plasma of radius $r_d = 5\,\mu m$ at density $\overline{N_e}$ up to $N_{cr}/2$. For $\overline{N_e} > N_{cr}/2$, the two approaches diverge and become more sensitive to the size parameter $k r_d$. Figure S1 shows this comparison, with the full Mie calculation of the effective refractive index contribution (change in total effective index) of each site $\Delta n_{eff,site}^{Mie} = (n_{eff}^{Mie} - 1)/N_{sc} V$ for plasma sites of various sizes as a function of $f = \overline{N_e}/N_{cr}$, along with the refractive index contribution per site used in simulations, $\Delta n_{eff,site}^{approx} = (n_{eff} - 1)/N_{sc} V = -(\overline{N_e}/2N_{cr})$. The actual change in local refractive index at each $\Delta n_{site} = (1 - \overline{N_e}/N_{cr})^{1/2} - 1$, is also shown.

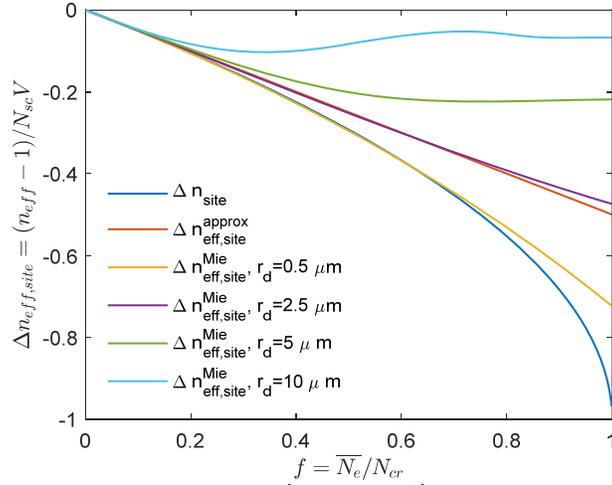

**Figure S1.** Effective index contribution per site $\Delta n_{eff,site}^{Mie} = (n_{eff}^{Mie} - 1)/N_{sc}V$ for various breakdown sizes as a function of $\overline{N_e}/N_{cr}$ at a site. The effective index contribution per site used in simulations $\Delta n_{eff,site}^{approx} = (n_{eff} - 1)/N_{sc}V = -(\overline{N_e}/2N_{cr})$ as well as the actual index change at each site $\Delta n_{site} = n - 1 = (1 - \overline{N_e}/N_{cr})^{1/2} - 1$ is shown for comparison.

At saturation, the growth of the avalanche is no longer driven by exponential growth of free electrons but rather by laser heating mediated by electron-ion collisions, accompanied by an expanding ionization shock front, which increases the breakdown size. Given the difficulty of fulling calculating evolving breakdown sizes, the resultant size dependent index of refraction, and increased scattering and the increase in attenuation, we assume saturation at full ionization with no further outward plasma growth, which would tend to underestimate losses and scattering for the back half of the pulse.

## 4. Temperature equilibrium

Under thermal quasi-equilibrium, the electron velocity distribution is given by $f(v) = \alpha v^2 \exp(-mv^2/2k_B T)$ for normalization constant $\alpha$. In order to achieve this state during laser heating, the electron-electron collision frequency $\nu_{ee}$ should be greater than the electron heating rate $\tau_h^{-1} \sim (K_e/(2U_p \nu_{en}))^{-1}$ for average electron kinetic energy $K_e = \frac{1}{2}m\langle v^2 \rangle$, where $\tau_h \sim 1$ ps is the heating timescale under our conditions. The electron-electron collision rate is $\nu_{ee}[\text{s}^{-1}] \approx 2.9 \times 10^{-6} \ln\Lambda\, N_e/T^{3/2}$, where $\ln\Lambda \approx 23 + \ln(T^{3/2} N_e^{-1/2})$ for $N_e$ in cm$^{-3}$ and $T$ in eV [3]. For $K_e \sim 10$-100 eV and $\ln\Lambda \sim 6-16$ under our simulation conditions, this gives $\nu_{ee} > \tau_h^{-1}$ only for $\overline{N_e} \gtrsim 10^{18}$ cm$^{-3}$. However, use of a thermal distribution in our calculations is still permissible over a much wider range of density because $\nu_{en} \gg \nu_{ee}$ during the fast rising portion of the breakdown. Momentum transfer collisions with neutrals impart an average energy of $2U_p$, but a single collision can add to (or deduct from) an electron's energy over a wide range, and thus have a "thermalizing" effect similar to electron-electron collisions even though there is average net energy gain. This is evident in calculations [4] and measurements [5,6] of inverse bremsstrahlung in fully ionized plasmas giving super Maxwellian distributions, namely $f_n(v) = \alpha_n v^2 [\exp(-mv^2/2k_B T)]^n$ for integer $n \geq 1$. Laser-driven collisions lead to large electron velocity variations and smoothing of the velocity distribution function. This is particularly relevant for intense LWIR beams, for which the peak laser-driven energy ($2U_p$) can be on the order of 100 eV (for a 5 TW/cm², λ=10 μm pulse), comparable to our simulated temperatures (~50-200 eV).

Below in Fig. S2, we plot ionization growth rates $v_{is}$ calculated from several different electron velocity distributions: a Gaussian distribution centered around a nonzero energy, and a super Maxwellian of order $n = 2$. As in the main text, $v_{is}$ is found for each distribution by finding the central energy/super-Maxwellian temperature where heating and loss rates balance as a function of intensity. These changes in distribution shape do not noticeably shift rates calculated using a standard Maxwellian ($n = 1$).

Our approach assumes that only elastic momentum transfer collisions contribute to electron heating, but an electron which undergoes an inelastic collision will also dephase with respect to the laser field in the same way as in the collisional heating process for elastic collisions. Correctly incorporating such additional heating is difficult [7], and suggests a Boltzmann approach. However, if we assume that including inelastic collisions effectively doubles the collisional heating rate, the ionization rate is moderately increased (× 2.5) at low intensities, with a smaller relative change at higher intensities.

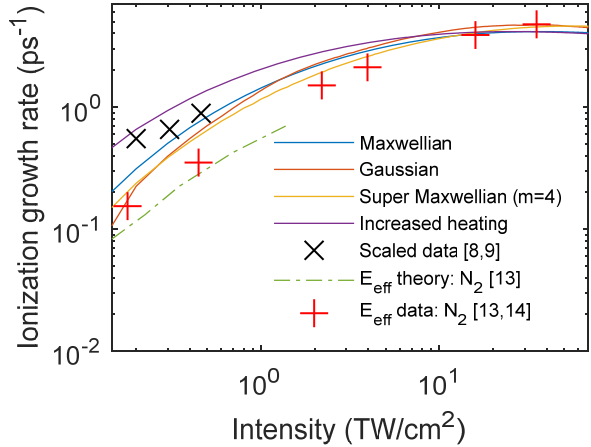

**Figure S2.** Quasi-equilibrium ionization growth rate $v_{is}$ vs. laser peak intensity (for λ=10 μm) for several electron velocity distributions. Purple curve: rough estimate of effect on $v_{is}$ of inelastic collisions. Also shown are scaled theoretical and experimental results, as described below.

As discussed in the main text, data from scaled experiments can also be compared to our calculated $v_{is}$. For $\omega \gg v_{en}$, the heating rate is expected to scale directly with the ponderomotive energy $U_p \propto I\lambda^2$, such that avalanches at λ=4 μm can be compared with those at λ=10 μm driven at a factor of $(10/4)^2 = 6.25$ lower intensity. A recent experiment directly measured the growth rate $v_{is} = 0.55$ ps$^{-1}$ for λ=4 μm at 1.3 TW/cm² [8], which is scaled and plotted as the lowest × point in Fig. S2. The remaining two × points are calculated from backscatter-based breakdown timing measurements in [9], assuming that the same growth $n_f/n_0 = \exp(0.55\tau(ps))$ is required to reach detection for total pulse length $\tau = 50$ ps. Due to the difficulty in comparing different experimental setups and backscatter geometries, these points are more uncertain, but still show that the temperature model predicts the general trend. For DC breakdowns, comparisons can be made using two experiments that have the same effective electric field, namely, $E_\omega v_{en}/\omega N_n = E_{DC}/N_n$ for electric field $E_\omega$ of frequency $\omega$ scaled by the neutral density $N_n$ in the experiment, where there are different conventions for whether the peak or rms field should be used [10-12]; here we scale by the peak field. Applying this scaling to DC breakdown theory [13] and experiments [13,14] in pure $N_2$ gives the remaining data in the plot, and again shows agreement within a factor of 2.

Because electron-neutral interactions dominate during the fast rising portion of the breakdown, and the neutral density is uniform across a breakdown site, our model predicts a very weak temperature gradient. Once a local breakdown nears saturation, however, the rates of electron-electron and electron-ion collisions rapidly increase as neutrals deplete. As such, strong temperature gradients will emerge, necessitating an additional level of complexity for modeling.

## 5. Aerosol breakdown rates and density scan

As discussed in the main text, to estimate aerosol-initiated avalanche growth, we use an ionization rate of $\nu_{is} = 10 \times I[\text{TW/cm}^2]$ ps$^{-1}$ calculated for fused silica breakdowns at $\lambda = 1$ μm [15] as representative. Using a rate directly proportional to intensity, rather than tracking temperature evolution, is permissible since the pulse duration is long compared to heating times, $\tau \gg \tau_h \sim K_e/2U_p\nu_{en}$, discussed in Sec. 4 above. To model ionization of sub-wavelength-sized particles, we apply $I \rightarrow 4I$ to account for $2\times$ electric field enhancement near a dielectric sphere with a dielectric constant $\varepsilon \sim 4$. This approach then predicts that at $I = 0.5$ TW/cm$^2$, growth will proceed from an initial electron seed in a 0.2 μm diameter aerosol to full single ionization ($10^8$ electrons) in $\Delta t \sim \ln(10^8)\,\nu_{is}^{-1} \sim 1$ ps. This is consistent with aerosols avalanching to saturation during the rise of the pulse and then avalanching more slowly as plasma expands into the surrounding air. While we have considered wavelength independent avalanche in aerosols for $\omega/\nu_{en} \ll 1$, Brunel or "vacuum" heating [16] at the particle-air interface may introduce additional wavelength-dependence.

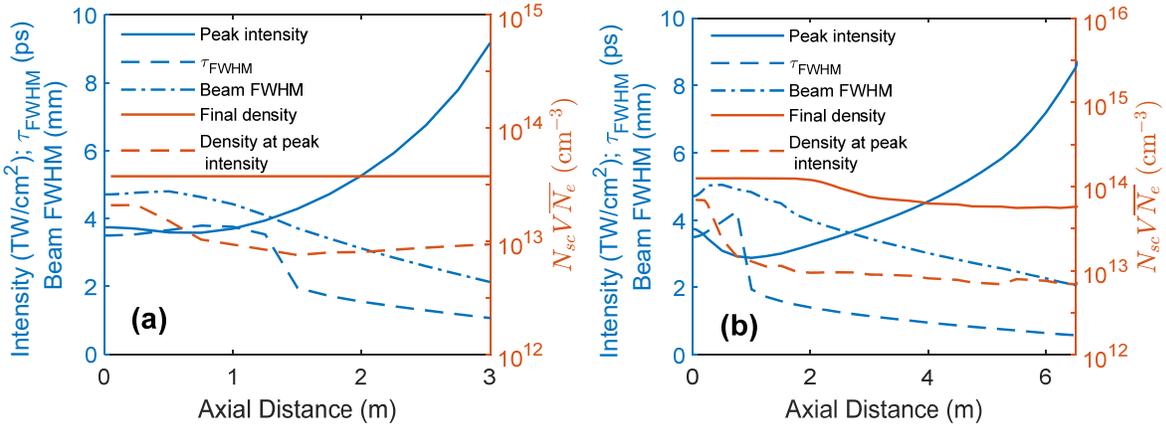

**Figure S3.** Simulation including aerosol-enhanced avalanche for initial aerosol densities (a) $N_{sc} = 3 \times 10^3$ cm$^{-3}$ and (b) $N_{sc} = 10^4$ cm$^{-3}$.

Additional simulation runs with initial aerosol densities less than $N_{sc} \sim 2 \times 10^4$ cm$^{-3}$ are shown below in Figure S3. While the run at (a) $N_{sc} = 3 \times 10^3$ cm$^{-3}$ shows slightly slower self-focusing than the case with no aerosols, the generated density is still insufficient to fully arrest pulse collapse, leading to a rapid increase in intensity as the pulse self-focuses towards the end of the run. For (b) $N_{sc} = 10^4$ cm$^{-3}$, the intensity decreases slightly before increasing again and trending upwards at the end of the 6.5 m simulation window. Further, since (a) and (b) show continued self-focusing (arrested only by tunneling ionization), the generated avalanches always exceed $\overline{N_e} > N_{cr}/2$ by the end of the pulse, limiting the validity of the propagation simulations.

## 6. Propagation/avalanche simulations

Here we present further details of our propagation/avalanche simulations. Our simulations use a 2D UPPE algorithm [17] to simulate MIR laser pulse propagation through air. UPPE ('Unidirectional Pulse Propagation Equation') is a system of ordinary differential equations (ODEs) of the form

$$\frac{\partial}{\partial z} A_{k_\perp}(\omega, z) = i Q_{k_\perp}(\omega) 2\pi P_{k_\perp}(\omega, z) e^{-i\left(k_z - \frac{\omega}{v_g}\right)z} . \tag{S.7}$$

Here, $A$ is related to the optical field $E$ by $E = A e^{i k_z z}$. The ODEs are indexed by $k_\perp$, the spectrum of radial spatial frequencies. The system is solved by a GPU implementation of the MATLAB ODE45 function. $P_{k_\perp}(\omega, z)$ is the nonlinear polarization of the medium, which in our simulation includes Kerr self-focusing, the molecular rotational response, and the plasma response. Our particular implementation of UPPE is called YAPPE ('Yet Another Pulse Propagation Effort'). The plasma response, which is the primary focus of our study, includes refraction and associated losses. The refractive response can be written as a polarization as follows:

$$P_{k_\perp}^{plasma}(\omega, z) = -\frac{4\pi e^2}{m_e \omega_c^2} \{N_{e,eff} E\}_{\omega, k_\perp} \tag{S.8}$$

for central frequency $\omega_c = 1.848 \times 10^{14}\ s^{-1}$ (corresponding to $\lambda = 10.2$ μm), plasma density $N_{e,eff}$, and optical field $E$, and where $\{\ \}_{\omega, k_\perp}$ denotes the following sequence of operations: the bracket contents (in Eq. S.8, the product $N_{e,eff}(r, z, t) E(r, z, t)$) are first computed in spatiotemporal space, then Fourier transformed $t \to \omega$, and then Hankel transformed $r \to k_\perp$. The plasma response is taken to be non-dispersive, using only the central frequency $\omega_c$ instead of all frequency components. Given that we simulate a 3.5ps duration pulse, the characteristic plasma dispersion length is long enough that this is a reasonable assumption. Tunneling/MPI losses are also factored into the simulation with the following imaginary polarization:

$$P_{k_\perp}^{ionloss}(\omega, z) = i \frac{n_0}{k} \left\{ \frac{(\partial_t N_e^{field}) U_I}{I} \right\}_{\omega, k_\perp} \tag{S.9}$$

for ionization energy $U_I$ and optical intensity $I$. $N_e^{field}$ is the electron density contributed by tunneling ionization/MPI; this equation does not include electron density yield from avalanche. There are also losses from plasma heating, which are explained after the avalanche model below.

During breakdown, electron temperature is tracked as

$$\frac{d k_B T}{dt} = \frac{2}{3}\left(2 U_p \nu_{en} + \nu_i U_p - \Sigma_l \nu_l \chi_l\right) - \nu_i T , \tag{S.10}$$

for ponderomotive energy $U_p \cong 0.93\, I\,[\text{TW/cm}^2](\lambda[\mu m])^2$ eV, electron neutral collision rate $\nu_{en}$ and ionization losses of energy $\chi_l$ and excitation rate $\nu_l$ in N₂ and O₂. All of these rates are self-consistently calculated as $\nu_l = \langle \sigma_l v \rangle N_n = \langle \sigma_l v \rangle (N_{n0} - N_e)$, where $N_n$ is the local neutral molecule density depleted from its initial value $N_{n0}$ as the electron density $N_e$ increases, and $\langle \sigma_i v \rangle = \int_0^\infty dv\, f(v) \sigma_i(v) v$ for electron velocity distribution $f(v) = \alpha v^2 \exp(-mv^2/2k_B T)$ at temperature $T$, with normalization constant $\alpha$. Values of relevant cross sections $\sigma_l$ are drawn from [18,19] as a function of electron energy (velocity). Equation (S.10) is implemented by using the data underlying Fig. 1(a) of the main text as a 'look-up table' for the heating rate $2U_p(2\nu_{en} + \nu_i)/3$, loss rate $2(\Sigma_l \nu_l \chi_l)/3 + \nu_i T$, and ionization rate $\nu_i$ as functions of temperature. Equation (S.10) is then solved using a first-order Forward Euler scheme with an initial temperature $T(t=0) = U_p$.

The number density of avalanche breakdown sites $N_{sc}$ is tracked from its initial value as

$$\frac{dN_{sc}}{dt} = \sum_i \nu_{i,MPI}(I)\, N_i \quad (\text{S. 11})$$

for tunneling/MPI ionization rate $\nu_{i,MPI}(I)$ of neutral species $i$ (taken to be O₂ and the $\chi_p \sim 6$ eV contaminant [9]) of density $N_i = N_{i,0} - N_{i,sc}$, accounting for prior ionization. We assume that these electrons, released into the laser field, have an initial effective temperature $U_p$; for a rapid increase in the density of tunneling/MPI generated electrons, this can lead to a lower effective temperature than in the purely collisionally-driven case. Given the time-dependent temperature, the evolution of the effective electron density $N_{e,eff} = N_{sc} V \overline{N_e}$, which we track as a single quantity, is given by

$$\frac{dN_{e,eff}}{dt} = N_{sc} \nu_i V \overline{N_e} + V \overline{N_e} \frac{dN_{sc}}{dt} \rightarrow N_{sc} \nu_i V \overline{N_e} + \frac{dN_{sc}}{dt} = \nu_i N_{e,eff} + \frac{dN_{sc}}{dt} \quad (\text{S. 12})$$

where the quantities not defined in this supplement are defined in the main text. To model what is really a spatially inhomogeneous process, we must apply separate physical considerations to each term of Eq. (S.12). The first term, $N_{sc} \nu_i V \overline{N_e}$, tracks the rate of avalanche ionization growth at existing breakdown sites. In the second term, putting $V\overline{N_e} \to 1$, so that $V\overline{N_e}\, dN_{sc}/dt \to dN_{sc}/dt$, enables tracking of only the contribution of new avalanche seed sites in neutral gas.

The effective transient refractive index is then given by

$$n_{eff} = 1 - \frac{N_{e,eff}}{2N_{cr}} = 1 - \frac{N_{sc} V \overline{N_e}}{2N_{cr}} \quad . \quad (\text{S. 13})$$

We note that saturation and depletion can occur either through volume average full single ionization of the gas, $N_{no} = N_{e,eff} = N_{sc} V \overline{N_e}$, or through full single ionization at a single avalanche site, $N_{no} = \overline{N_e}$. Thus, we take the transient residual neutral density during ionization to be

$$N_n = \min(N_{no} - N_{e,eff}, N_{no} - \overline{N_e}), \tag{S.14}$$

for $\overline{N_e} = N_{e,eff}/(N_{sc}(4\pi r_d^3/3))$. Avalanches seeded by aerosols were assumed to start with one electron per aerosol, either provided by existing static charge or by ionization early in the pulse. As discussed in Sec. 5, electron number growth at each aerosol follows $n_e(t) = n_{e,0}e^{\nu_{is}t}$ for $\nu_{is} = 40 \times I[\text{TW/cm}^2]\text{ ps}^{-1}$, which continues until the $\sim 10^8$ atoms in a $\sim 0.2$ µm diameter particle are singly ionized. To make our simulation tractable while still preserving the physics relevant to propagation, we ignore the plasma dynamics of aerosol explosion into the surrounding air: $n_e(t)$ is combined with avalanche in the surrounding air computed with Eqs. (S.10) − (S.12), forming $\overline{N_e} \to \overline{N_e} + n_e(4\pi r_d^3/3)^{-1}$. Since aerosols avalanche (and saturate) so much more quickly than breakdowns seeded by single electron seeds, they can be considered the dominant source of avalanche when they are present.

Absorption of laser energy density due to laser heating is included in the YAPPE propagation simulations as a laser energy loss rate per unit volume

$$W_{abs} = \frac{2}{3}\left(2U_p\nu_{en} + \nu_i U_p\right)N_{e,eff}. \tag{S.15}$$

This energy loss rate is also modelled as a complex polarization:

$$P_{k_\perp}^{abs}(\omega, z) = i\frac{n_0}{k}\left\{\frac{W_{abs}}{I}E\right\}_{\omega, k_\perp} \tag{S.16}$$

For plasma generation below saturation, we neglect scattering losses, as discussed in Sec. 2 above.